\documentclass[journal,twoside]{IEEEtran}

\usepackage{etoolbox}
\makeatletter
\@ifundefined{color@begingroup}%
{\let\color@begingroup\relax
\let\color@endgroup\relax}{}%
\def\fix@ieeecolor@hbox#1{%
\hbox{\color@begingroup#1\color@endgroup}}
\patchcmd\@makecaption{\hbox}{\fix@ieeecolor@hbox}{}{\FAILED}
\patchcmd\@makecaption{\hbox}{\fix@ieeecolor@hbox}{}{\FAILED}

\usepackage{cite}
\usepackage{amsmath,amssymb,amsfonts}
\usepackage{algorithmic}
\usepackage{algorithm}
\usepackage{graphicx}
\usepackage{textcomp}
\usepackage{multirow}
\usepackage{tabularx}
\def\BibTeX{{\rm B\kern-.05em{\sc i\kern-.025em b}\kern-.08em
    T\kern-.1667em\lower.7ex\hbox{E}\kern-.125emX}}

\begin{document}
\title{Enhancing Subject-Independent Accuracy in fNIRS-based Brain-Computer Interfaces with Optimized Channel Selection}

\author{
    Yuxin Li, Hao Fang, Wen Liu, Chuantong Cheng, Hongda Chen
    \thanks{Yuxin Li is with the School of Mechanical Engineering, Xi’an Jiaotong University, Xi'an, China.}
    \thanks{Hao Fang is with the SMART Group, Institute for Imaging, Data and Communications, School of Engineering, The University of Edinburgh, Edinburgh, UK.}
    \thanks{Wen Liu is with the Department of Electrical and Electronic Engineering, Xi'an Jiaotong-Liverpool University, Suzhou, China.}
    \thanks{Chuantong Cheng and Hongda Chen are with the State Key Laboratory on Integrated Optoelectronics, Institute of Semiconductors, Chinese Academy of Sciences, Beijing, China.}
    \thanks{This work was supported by the National Key R\&D Program of China (Grant Nos. 2021YFB3601201). Correspondence author: Wen Liu and Chuantong Cheng; Email: Wen.Liu@xjtlu.edu.cn; chengchuantong@semi.ac.cn}
}


    

\maketitle


\begin{abstract}

Achieving high subject-independent accuracy in functional near-infrared spectroscopy (fNIRS)-based brain-computer interfaces (BCIs) remains a challenge, particularly when minimizing the number of channels. This study proposes a novel feature extraction scheme and a Pearson correlation-based channel selection algorithm to enhance classification accuracy while reducing hardware complexity. Using an open-access fNIRS dataset, our method improved average accuracy by 28.09\% compared to existing approaches, achieving a peak subject-independent accuracy of 95.98\% with only two channels. These results demonstrate the potential of our optimized feature extraction and channel selection methods for developing efficient, subject-independent fNIRS-based BCI systems.

\end{abstract}


\section{Introduction}
\label{sec:introduction}
\IEEEPARstart{T}he advancements in the field of brain-computer interfaces (BCI) have fueled scholarly enthusiasm for the study of cerebral activities. Studying cerebral activity enables interaction between the human cerebral and external devices, with significant contributions to fields such as rehabilitation medicine, psychology, and military. Commonly employed methodologies for evaluating cerebral activity across distinct task states encompass functional near-infrared spectroscopy (fNIRS), functional magnetic resonance imaging (fMRI), and electroencephalography(EEG)\cite{ref36}. Notably, fNIRS boasts distinct advantages over alternative neuroimaging modalities, primarily manifesting in its inherent portability, safety profile, and capacity to withstand environmental electrical noise. Additionally, fNIRS exhibits heightened resilience to motion artifacts and electromyographic (EMG) interferences when juxtaposed with methodologies such as EEG \cite{ref1}. Conversely, while fMRI has been prolific in elucidating the blood oxygenation level-dependent (BOLD) signal, particularly during visual stimuli, fNIRS has unequivocally established correlations between alterations in blood hemoglobin concentration and the presence or absence of external stimulation\cite{ref1,ref8,ref9}. This positioning of fNIRS as a compelling choice for scrutinizing diverse task states within the intricacies of the human brain is thus underscored.

fNIRS devices have a simple structure, and the measurement principle is not overly complex. An fNIRS device typically includes multiple light sources and detectors, placed on the scalp. The emitted near-infrared light passes through the scalp and brain tissue before being captured by the detectors. By examining how light is absorbed and scattered, it becomes possible to deduce alterations in oxygen levels in different brain regions, thus enabling the study of brain function. fNIRS employs near-infrared light, situated between visible and infrared wavelengths, to measure changes in the concentration of hemoglobin and oxyhemoglobin in brain tissue\cite{ref31}. When specific brain regions are active, blood flow increases, leading to variations in oxygen levels\cite{ref35}. fNIRS calculates these oxygenation changes by analyzing differences in the intensity of incoming and outgoing light. 

The pivotal pursuit of discerning neural activation patterns in the human brain during diverse task states constitutes a fundamental thrust in the practical advancement of BCIs. By measuring hemoglobin concentration changes, fNIRS provides insights into neural activities, positioning it as a promising tool for various BCI tasks, including cognitive, motor, visual, and visuo-motor functions. \cite{ref2,ref3,ref4,ref5,ref6,ref7,ref15}. In 2009, Huppert, T. J., et al\cite{ref16}, utilized Support Vector Machines to classify brain activity during the execution of working memory tasks, achieving an accuracy rate of approximately 85\%. In 2017, Hong, K. S. \cite{ref17}, employed Convolutional Neural Networks (CNN) for classifying experiments that involved controlling external devices through brain activity, achieving an accuracy rate exceeding 90\%. In 2018, Santosa, H. \cite{ref18}, used Linear Discriminant Analysis to classify subjects based on different emotional states, achieving an accuracy rate of over 80\%. While numerous studies\cite{ref13} have delved into enhancing the performance of fNIRS-based BCIs, the dual challenge of subject-independent accuracy and online classification remains largely unaddressed. Most existing methodologies either focus on one aspect, sacrificing the other, or require a large number of subjects and extensive individual calibration sessions, which are not feasible for real-world applications\cite{ref10,ref11,ref14}.

Depending on specific research questions and interests, different types of tasks and classifiers can be selected for fNIRS users. However, there are still some challenges in this field: One of the paramount challenges in the development and deployment of fNIRS-based BCI devices is achieving high subject-independent accuracy. This refers to the algorithm's ability to maintain consistent classification accuracy across different individuals, ensuring that its recognition capabilities are not confined to subjects-dependent but can be generalized across a diverse population\cite{ref25}. Such generalization is crucial, as individual differences in brain anatomy, physiology, and cognitive strategies can introduce variability in neural signals. Moreover, the demand for online classification, which involves real-time processing and interpretation of neural data, further complicates the BCI design.


Our study pioneers a novel optimization scheme specifically tailored for binary classification tasks to improve the accuracy of classification. This scheme aims to discover the characteristics of fNIRS data and develop a feature extraction scheme that aligns with hemodynamics for training machine learning classifiers. These features include statistical features, time-domain features, frequency-domain features, and principal component features. Extracting statistical features for different brain regions can construct more accurate classification models. 
The strength of our scheme lies in its holistic approach, amalgamating diverse feature sets that collectively offer a rich, multi-faceted view of the data. This ensures that our models are informed by  temporal, statistical, and spectral variations of the signals, positioning them for superior performance in classification tasks.

When conducting channel selection, confronted with the complexity of a vast number of channels, we employ a Pearson correlation-based approach to construct an adjacency matrix. This enables us to quantify the information content within each channel, with the objective of minimizing information loss while reducing the number of channels. This reduction is aimed at enhancing classification accuracy across distinct task states. Central to our approach is a strategic reduction in hardware components, ensuring their optimal positioning, coupled with an innovative feature extraction methodology that adeptly captures the unique attributes of fNIRS signals.


To validate the effectiveness of our scheme, we conducted experiments on a 52-channel binary classification fNIRS dataset related to mental arithmetic. It's noteworthy that our experimental setup maintained consistency with G. Bauernfe et al.\cite{ref12} in terms of utilizing the Linear Discriminant Analysis (LDA) algorithm and the same dataset. With this setup, the introduction of our feature extraction scheme led to a notable 28.09\% improvement in average accuracy. This improvement is the mean value derived from three sets of experiments using oxy-Hb data from three ROIs. Specifically, for ROI$_1$, ROI$_2$, and ROI$_3$, the accuracy enhancements were 25.44\%, 29.56\%, and 29.28\%, respectively. Furthermore, our channel selection algorithm enabled us to achieve a peak subject-independent accuracy of 95.98\% using only two channels: channel 26 in the anterior prefrontal cortex (aPFC) and channel 43 in the right dorsolateral prefrontal cortex (r.DLPFC) on the same dataset.

The rest of this paper is organized as follows. The second section presents the methods used in our scheme. The third section describes the experimental setup of our research, including the introduction of the dataset used, experimental parameter settings, and evaluation metrics. The fourth section presents the experimental results. The fifth section is the conclusion, summarizing the advantages of our scheme and future prospects.

\section{Methods}
\subsection{Feature Extraction}
Feature extraction stands as a cornerstone in the domain of machine learning and signal processing, particularly for complex datasets like those derived from fNIRS\cite{ref30}. 
fNIRS dataset can be noisy due to instruments and the environment. To make sense of this data, researchers extract useful features and use machine learning to classify different brain states \cite{ref26}. This process not only aids in enhancing the interpretability of the data but also plays a pivotal role in improving the efficiency and performance of subsequent analytical tasks.

\begin{figure*}[htbp]
	\centerline{\includegraphics[scale=0.65]{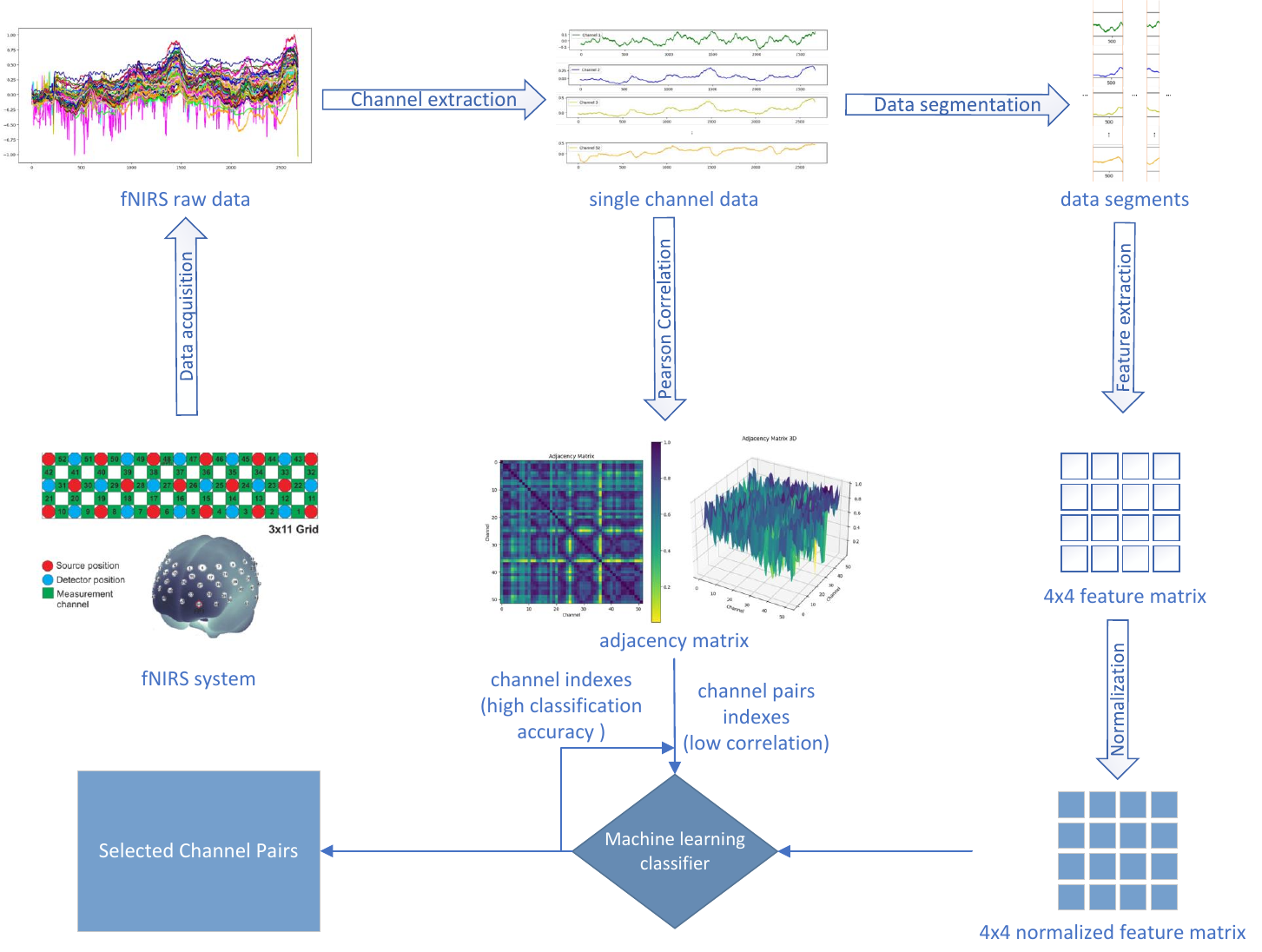}}
	\caption{Flow chart of our  methods.} 
	\label{flowchart}
\end{figure*}

\subsubsection{Statistical Features}
By computing the mean, maximum, minimum, and variance of the data, we succinctly capture its central tendency, range, and variability. The mean reflects the average blood oxygen level over time, while the maximum and minimum values help identify peaks and valleys in brain blood flow, indicating brain activity. Variance measures the instability in the hemodynamic response.
\subsubsection{Frequency-domain Features}
The Fourier Transform converts the time-domain data from fNIRS measurements into frequency-domain data, providing information about the different frequency components. It helps to detect changes in blood oxygen at different frequencies.
\subsubsection{Temporal Rate-of-change Features}
By calculating the slope and its associated mean, maximum, minimum, and variance, we gauge the dynamic changes and trends in the time domain. This reveals the rate at which neural activities change over time, highlighting rapid fluctuations or steady states. The slope measures the rate of change in the blood oxygen response, which is important for capturing dynamic changes in brain activity.
\subsubsection{Principal Component Statistical Features}
By leveraging Principal Component Analysis (PCA), we distill the data into its primary modes of variation. This helps to understand the relationships between brain activity characteristics and which characteristics are most important for a particular task or condition.

\begin{equation}	
	\begin{split}
    feature\_matrix(raw\_data\_segment) = \\
    \left[ \begin{array}{cccc}
    mean & max & min & var\\
    pca_{\text{mean}} & pca_{\text{max}} & pca_{\text{min}} & pca_{\text{var}}\\
    slope_{\text{mean}} & slope_{\text{max}} & slope_{\text{min}} & slope_{\text{var}}\\
    fft_{\text{mean}} & fft_{\text{max}} & fft_{\text{min}} & fft_{\text{var}}\\
    \end{array}
    \right ],
    \end{split}
\end{equation}

The equation provided offers a structured representation of a feature matrix derived from a segment of single-channel raw fNIRS data, denoted as $raw\_data\_segment$.This segment captures a specific temporal window of neural activity. Collectively, this feature matrix ensures that subsequent analyses, especially machine learning models, have access to a rich set of features that comprehensively capture the nuances of the underlying neural activity.

\subsection{Normalization}


After the feature extraction process, the importance of normalization becomes even more pronounced. The fNIRS data usually have different sizes and magnitudes, which if not normalized may lead to some features being overweighted in classification, affecting the classification results. Scaling the data into [0, 1] ensures that all features have similar scales. The formula for Min-Max normalization is:

\begin{equation}
X_{\text{norm}} = \frac{X - X_{\text{min}}}{X_{\text{max}} - X_{\text{min}}}
\end{equation}

Where $X_{\text{norm}}$ is the normalized value, $X$ is the original value, and $X_{\text{max}}$ and $X_{\text{min}}$ are the minimum and maximum values of the feature, respectively.

By applying normalization post-feature extraction, we ensure that these diverse features are harmonized in scale, setting the stage for machine learning models to discern and capitalize on the underlying patterns in the data more effectively.

\subsection{Machine Learning}

When tackling binary classification problems using fNIRS data, the selection of the analytical approach is crucial. While deep learning has achieved remarkable success in various areas, especially where large datasets are available, there are compelling reasons to choose traditional machine learning techniques for fNIRS:

\subsubsection{Data Volume Requirements}
Given the complexity and cost associated with many fNIRS experiments, the available data might be limited. Traditional machine learning methods, such as Support Vector Machines (SVM) or Random Forests, often perform well with smaller datasets.

\subsubsection{Computational Complexity}
Machine learning methods are generally more efficient and don't require specialized hardware like GPUs, More responsive to the needs of fNIRS online categorization.

\subsubsection{Model Interpretability}
Models like decision trees or linear regression in machine learning offer better interpretability. This is vital in fields like medicine or biology, where understanding the decision-making process of a model can be crucial. Deep learning models, in contrast, often act as "black boxes."

\subsubsection{Feature Engineering}
In machine learning, researchers can use domain knowledge for feature engineering, potentially improving model performance. For instance, We can use our well-designed feature extraction scheme suitable for fNIRS data classification in machine learning methods. Deep learning models, on the other hand, typically learn features automatically, which might not always be optimal.

In summary, the specific challenges and needs of fNIRS-based binary classification make traditional machine learning methods a more appropriate choice\cite{ref33}. In our experiments, we explored a diverse set of machine learning algorithms, including Support Vector Machine (SVM), Logistic Regression, Decision Tree, K-Nearest Neighbors (KNeighbors), Gaussian Naive Bayes (GaussianNB), Linear Discriminant Analysis (LDA), Multi-layer Perceptron Classifier (MLPClassifier), and Stochastic Gradient Descent (SGD).

Through rigorous testing and comparison, we identified that the MLP, LDA, and SVM consistently outperformed the others, emerging as the top three algorithms best suited for our fNIRS-based binary classification tasks. The architecture and principles of the above machine learning algorithm are:

\textbf{Multilayer Perceptron Classifier (MLP Classifier):}
The MLP Classifier constructs a multilayer neural network after extracting the fNIRS data features, which usually includes an input layer, one or more hidden layers, and an output layer\cite{ref27}. Each neuron receives inputs from the previous layer, weights and sums the inputs, and introduces a nonlinear transformation through an activation function. The weights and bias parameters of the neural network need to be initialized and adjusted to minimize the loss function by the backpropagation algorithm during the training process. The training process is designed to enable the neural network to automatically learn complex relationships between features from fNIRS data in order to accurately classify different task states or brain activity patterns. In this way, the MLP Classifier can help researchers mine useful information from fNIRS data for efficient brain activity classification.

\textbf{Linear Discriminant Analysis (LDA):}
LDA operates by linearly transforming data into a lower-dimensional space, often along a single line or hyperplane. Its primary goal is to maximize inter-class distinctions while minimizing intra-class variations. This approach significantly improves classifier performance\cite{ref28}. In the context of fNIRS data, which often results in high-dimensional, multi-channel time-series datasets, LDA is employed to project this data into a lower-dimensional space. This, in turn, facilitates a more accurate classification of distinct task states.
Furthermore, LDA aims to optimize differences between different categories, making it particularly valuable for investigating disparities among various regions of the brain.

\textbf{Support Vector Machine (SVM):}
Since SVM works well in solving multidimensional small sample sizes and nonlinear classification problems, it is suitable for fNIRS signal classification. SVM tries to find an optimal hyperplane to separate different classes of features, which should be perturbed by the training set as little as possible.
Given a training set of n samples {(x$_i$,y$_i$)},i=1,2,... ,y$_i$ are the sample labels, and find an optimal hyperplane that satisfies the requirements by constructing an objective function, which is described by a linear equation as  :

\begin{equation}
    w^Tx+b=0
\end{equation}
$w$ is a nonzero normal vector perpendicular to the separating hyperplane, and $b$ is the intercept. $d=\frac{2}{||w||_2}$ is the interval we inscribed.

The kernel function used in this study is a Gaussian kernel function, which can be mapped with no linear dimension and more diverse decision boundaries\cite{ref19}.

\subsection{Channel Selection}
\begin{algorithm} 
    
	\caption{Channel Selection} 
	\label{alg4} 
	\begin{algorithmic}
  
		\REQUIRE N-channel fNIRS raw data
		\ENSURE Optimal channel combinations and their accuracies
		\STATE Construct Pearson correlation for N channels
		\STATE Build NxN adjacency matrix based on correlations
		\STATE Sort matrix elements in descending order
        \STATE Select channel combinations with Pearson correlations less than \(0.4\) (indicating weak or very weak correlations) and record their indices
        \STATE Split data based on trials
		\FOR{each channel}
		    \STATE Construct feature matrix using the feature selection scheme
		    \STATE Normalize the feature matrix
		    \STATE Obtain normalized dataset for the channel
		\ENDFOR
		\FOR{each normalized channel dataset}
		    \STATE Apply machine learning algorithm with 5-fold cross-validation
		    \STATE Record accuracy for the channel
		\ENDFOR
		\STATE Sort all accuracies in ascending order
		\STATE Record top 20\% channel indices and their accuracies
		\STATE Scan the recorded single channel indices and previously selected channel combinations
		\STATE Identify channel combinations present in both lists
		\FOR{each identified channel combination}
		    \STATE Apply machine learning algorithm with 5-fold cross-validation on the dataset corresponding to the channel combination
		    \STATE Record accuracy for the channel combination
		\ENDFOR
		\STATE Sort all combination accuracies in ascending order
		\STATE Output all channel combinations and their accuracies that meet the criteria
	\end{algorithmic} 
\end{algorithm}

Channel selection is a pivotal step in the processing and analysis of multi-channel fNIRS data. In the realm of brain-computer interfaces (BCIs), where the data is often high-dimensional due to the multitude of channels, selecting the most informative channels becomes imperative. By focusing on channels that capture the most discriminative information, the accuracy of subsequent classification tasks can be significantly improved. Redundant or noisy channels can introduce variability that might confound the classifiers, leading to sub-optimal performance.

Moreover, in BCIs, real-time response is often crucial, especially for applications like prosthetic control or real-time neurofeedback. Processing data from fewer, but more informative channels can expedite the classification process, ensuring timely responses. From a hardware perspective, every channel in a BCI system corresponds to sensors, transmitters, and associated electronics. By narrowing down to essential channels, the overall cost of the BCI device can be reduced. Additionally, fewer channels mean less data to process, store, and transmit, leading to computational savings and more efficient power utilization.

Given the importance of channel selection, a channel selection algorithm is meticulously designed to identify the most informative and distinct channel combinations from raw data derived from N channels of fNIRS. Our method employs Pearson's correlation coefficient, a statistical measure that quantifies the linear relationship between two variables. The formula for Pearson correlation coefficient, $r$, is given by:

\begin{equation}
r=\frac{\sum_{i=1}^{n}\left(x_{i}-\overline{x}\right)\left(y_{i}-\overline{y}\right)}{\sqrt{\sum_{i=1}^{n}\left(x_{i}-\overline{x}\right)^{2} \sum_{i=1}^{n}\left(y_{i} - \overline{y}\right)^{2}}}
\end{equation}

where $r$ is the Pearson correlation coefficient, quantifies the linear relationship between two samples of data. $x_{i}$ and $y_{i}$ represent individual data points of samples $x$ and $y$, while $\overline{x}$ and $\overline{y}$ denote their respective means. The total number of data points is represented by n. The value of $r$ can range from -1 to 1, with 1 indicating a perfect positive linear relationship, -1 indicating a perfect negative linear relationship, and 0 suggesting no linear relationship.

\begin{table}[h]
\centering
\caption{An Typical Interpretation of Pearson Correlation Coefficient}
\begin{tabular}{|c|c|}
\hline
\textbf{Absolute value of $r$} & \textbf{Interpretation} \\
\hline
0.8 - 1.0 & Extremely Strong Correlation \\
\hline
0.6 - 0.8 & Strong Correlation \\
\hline
0.4 - 0.6 & Moderate Correlation \\
\hline
0.2 - 0.4 & Weak Correlation \\
\hline
0.0 - 0.2 & Very Weak or No Correlation \\
\hline
\end{tabular}
\end{table}

\begin{figure}[htbp]
	\centerline{\includegraphics[scale=0.6]{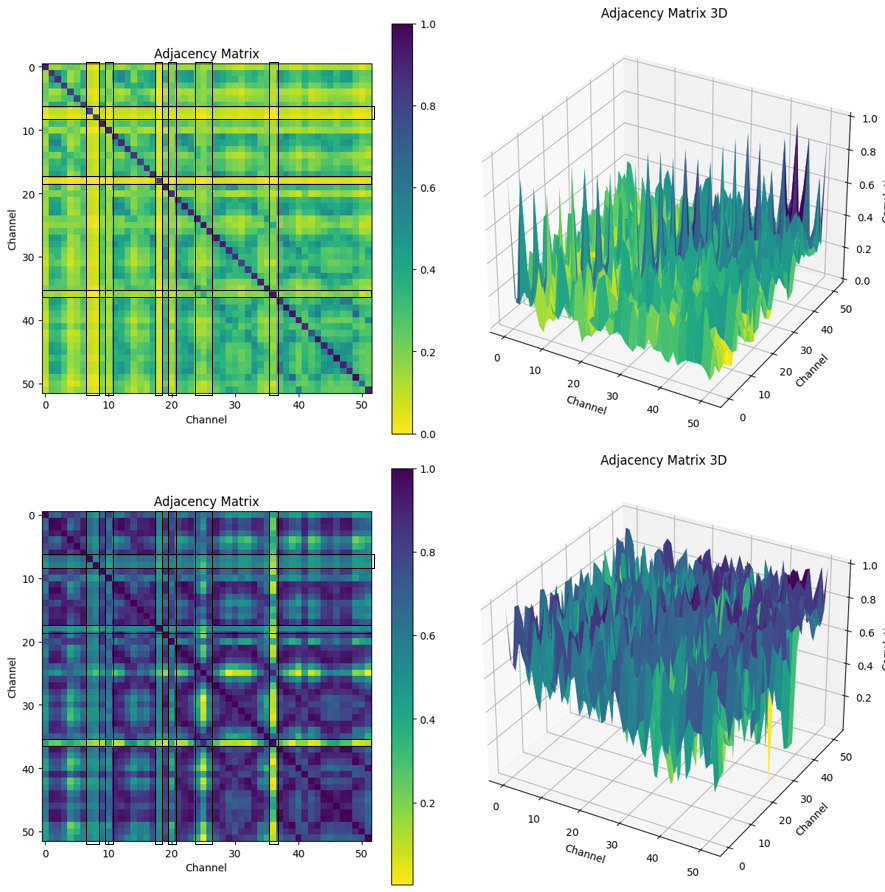}}
	\caption{Comparative visualization of adjacency matrices constructed using mutual information (top) and Pearson correlation (bottom). While the mutual information-based matrix exhibits a generally lighter colormap, the alignment of its bright and dark regions with those of the Pearson-based matrix is evident. Specifically, regions of low Pearson correlation align well with areas of low mutual information, and vice versa for high values, highlighting the consistency between the two methods.}
	\label{mutual_pearson}
\end{figure}

In our algorithm, we retain only those channel combinations with an absolute value of Pearson correlation below 0.4. This threshold indicates weak or very weak correlations, ensuring that the chosen channels are relatively independent and offer diverse information. To gain a more intuitive understanding of the correlation between different channels, we constructed visual 2D and 3D adjacency matrices based on the absolute values of the computed Pearson correlation, as illustrated in Fig.~\ref{mutual_pearson}.

At the same time, we also employed machine learning algorithms to obtain classification accuracies of models trained using the normalized feature matrices corresponding to each channel. We retained the channel identifiers for the top 20\% of channels based on their performance. After that, we sequentially retrieved channel combinations from those derived based on Pearson correlation that contained the aforementioned channel identifiers with high classification accuracy. We then used machine learning techniques to obtain the classification accuracies corresponding to these channel combinations. Ultimately, we ranked these channel combinations based on their classification accuracy from high to low. The combinations at the forefront are the dual-channel combinations selected by our channel selection algorithm.

\section{Experimental Setup}
\subsection{Dataset}
In this study, we utilized the 52-channel dataset constructed by Günther Bauernfeind et al. in 2014 for our experiments. This dataset encompasses recordings from eight subjects (three males, five females, with an average age of 26 and a standard deviation of 2.8 years) who exhibited contrasting hemodynamic response patterns during a mental arithmetic (MA) task. In this task, subjects were instructed to perform visually cued arithmetic calculations. For instance, they were tasked with continuously subtracting a single-digit number from a two-digit number (e.g., 97 - 4 = 93, 93 - 4 = 89, …) as swiftly as possible for a duration of 12 seconds, followed by a rest period of 28 seconds.


The data was recorded using a continuous-wave system (ETG-4000, Hitachi Medical Co., Japan) that measures changes in cerebral oxygenation and comprises 16 photodetectors and 17 photo emitters, resulting in a total of 52 channels. The sampling rate was set at 10Hz. The distance between the source and the detector was maintained at 3 cm. The lowest row of channels was aligned along the FP1–FP2 line of the international EEG 10–20 system, with channel 48 precisely positioned at FP1\cite{ref12}.

\begin{figure}[htbp]
	\centerline{\includegraphics[scale=0.30]{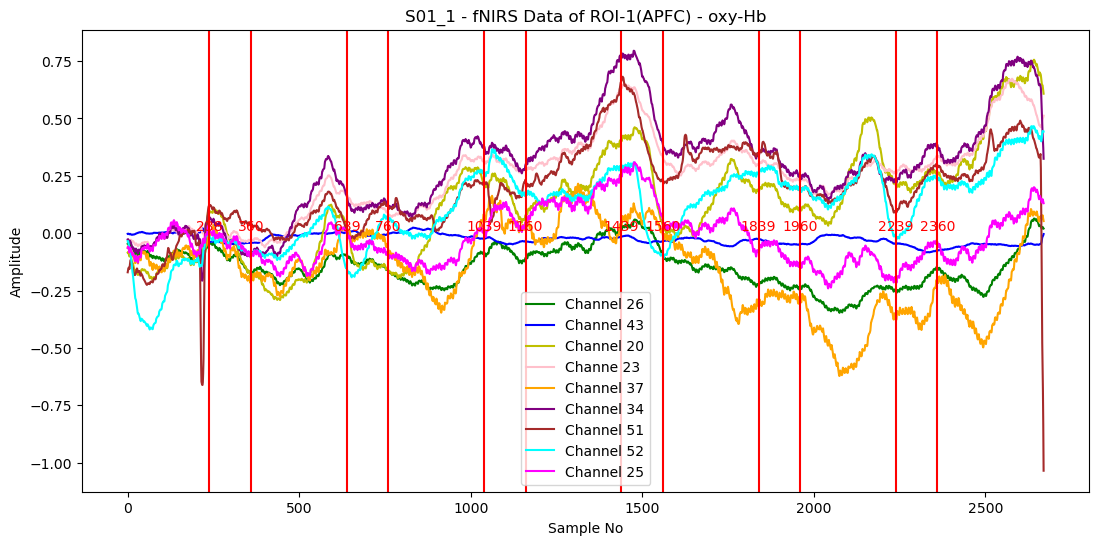}}
	\caption{Temporal variations in the concentration changes of oxy-hemoglobin across our selected 9 channels during a MA task for a single subject. Each channel's data is represented by a distinct color, with colors cycling through a predefined set for clarity. Vertical red lines indicate the start or end of an experimental trial, serving as markers to differentiate between rest periods and active MA phases. The x-axis denotes the sample number, serving as a proxy for time, while the y-axis indicates the amplitude of the oxy-hemoglobin concentration changes. Considering the time delay of the hemodynamic response, there are data rounding operations. The observed fluctuations provide insights into the cerebral hemodynamic responses associated with the cognitive demands of the task. }
	\label{raw_data_example}
\end{figure}

The MA dataset primarily consists of fNIRS raw data, positions of the trials, and labels. The fNIRS raw data reflects the changes in the concentrations of various substances (oxygen-hemoglobin, deoxygen-hemoglobin, total-Hb) in the brain during the experimental process. The positions of the trials, on the other hand, correspond to the start and end of the MA task. The labels illustrate the composition of the entire experimental process, where '1' indicates the MA task is in progress, and '2' signifies a resting state. For instance, Fig.~\ref{raw_data_example} depicts the variation in oxygen-hemoglobin concentration over time in nine channels for a participant throughout the experimental procedure. Within the figure, each curve corresponds to the concentration variation of the substance within a channel, while each vertical red line represents the position of a trial. To make the dataset suitable for machine learning classification, we preprocessed the data. This primarily involved segmenting the data based on the positions of the trials and aligning it with the labels.



\subsection{Experimental Setting}
After preprocessing the dataset, we first employed our feature extraction scheme to construct feature matrices. Then, these matrices undergo normalization, a crucial step to ensure all features have a consistent scale, which is pivotal for the effective application of many machine learning algorithms. Subsequently, we partitioned the dataset into 80\% for training and 20\% for testing. Finally, for training datasets generated based on different single channels or channel combinations, we trained and tested using various machine learning algorithms. 

Throughout the process, we set all random seeds to 56 to ensure the reproducibility of the experiments. Additionally, the kernel for our SVM classifier was set to 'linear'. For the MLP classifier, the size of the hidden layers was set to (10, 10, 10, 10), with a maximum iteration count of 3000.

\subsection{Evaluation Metrics}
We adopted various evaluation metrics to assess the performance of our scheme. These metrics include classification accuracy, sensitivity, specificity, Cohen's Kappa, F1 score, and ROC curve. Additionally, we employed 5-fold cross-validation as our training assessment strategy. 

\section{Experimental Evaluation and Results}

\subsection{Adjacency Matrix from Pearson correlation}
At the outset of our experiment, we extracted data from each of the 52 channels within the fNIRS dataset. This was followed by the computation of the Pearson correlation for every channel pairing, leading to the creation of a 52x52 Pearson correlation matrix. To further refine this matrix, we derived its absolute values, thereby constructing an adjacency matrix. Utilizing a sorting algorithm, we pinpointed channel combinations with absolute correlation values between 0 and 0.4, signifying weak or very weak correlations. 

In a bid to validate the relationship between Pearson correlation and the amount of shared information between two channels, we also constructed an adjacency matrix using mutual information. Upon visual inspection of the resulting data visualizations, we observed a striking similarity in the color distributions between the Pearson-based correlation map (c-map) and the mutual information-based c-map. Notably, even though the average values of mutual information were relatively smaller compared to the Pearson correlation values, the patterns of light and dark regions aligned well, suggesting a consistent relationship. Specifically, regions with low Pearson correlation corresponded seamlessly with areas of low mutual information, underscoring the reliability of our initial Pearson-based approach.

\begin{figure}[htbp]
	\centerline{\includegraphics[scale=0.35]{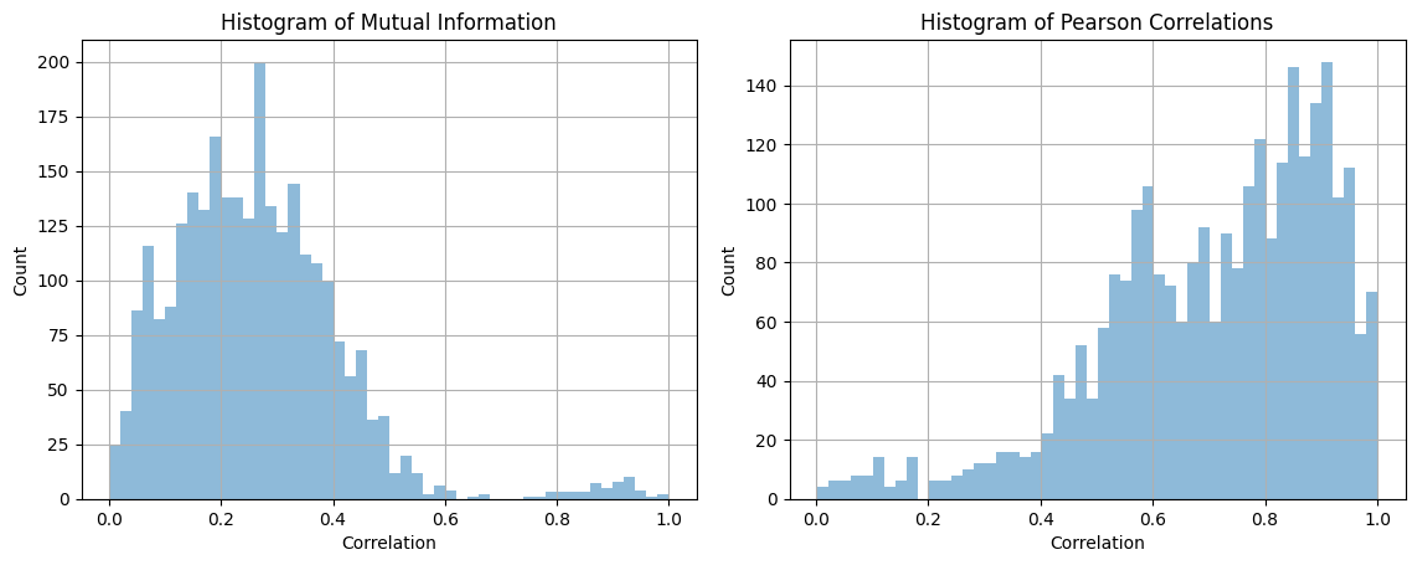}}
	\caption{Histogram comparison of values derived from adjacency matrices based on mutual information (left) and Pearson correlation (right). While the average values from the mutual information are notably smaller than those from the Pearson correlation, both histograms exhibit a congruent trend in counts as the values increase, underscoring the parallelism in their distribution patterns.}
	\label{histogram}
\end{figure}

\subsection{Feature Extraction, Normalization and Machine Learning}

Following the initial steps of our experiment, having segmented and labeled the raw data from each channel, the subsequent pivotal step was feature extraction. Using our crafted scheme, the raw data is transformed into feature data that can be used in a classifier.


Furthermore, we adopted two distinct evaluation strategies. In the "Subject-dependent classifications" machine learning algorithms were trained and tested independently on each subject's dataset, yielding individualized accuracy. Conversely, for "Subject-independent classifications," we amalgamated and shuffled the datasets of all subjects, aiming to assess the model's broader generalizability. 
This dual approach offered a holistic view of our model's performance, reinforcing the robustness and versatility of our experimental design.

\begin{figure*}[htbp]
	\centerline{\includegraphics[scale=0.62]{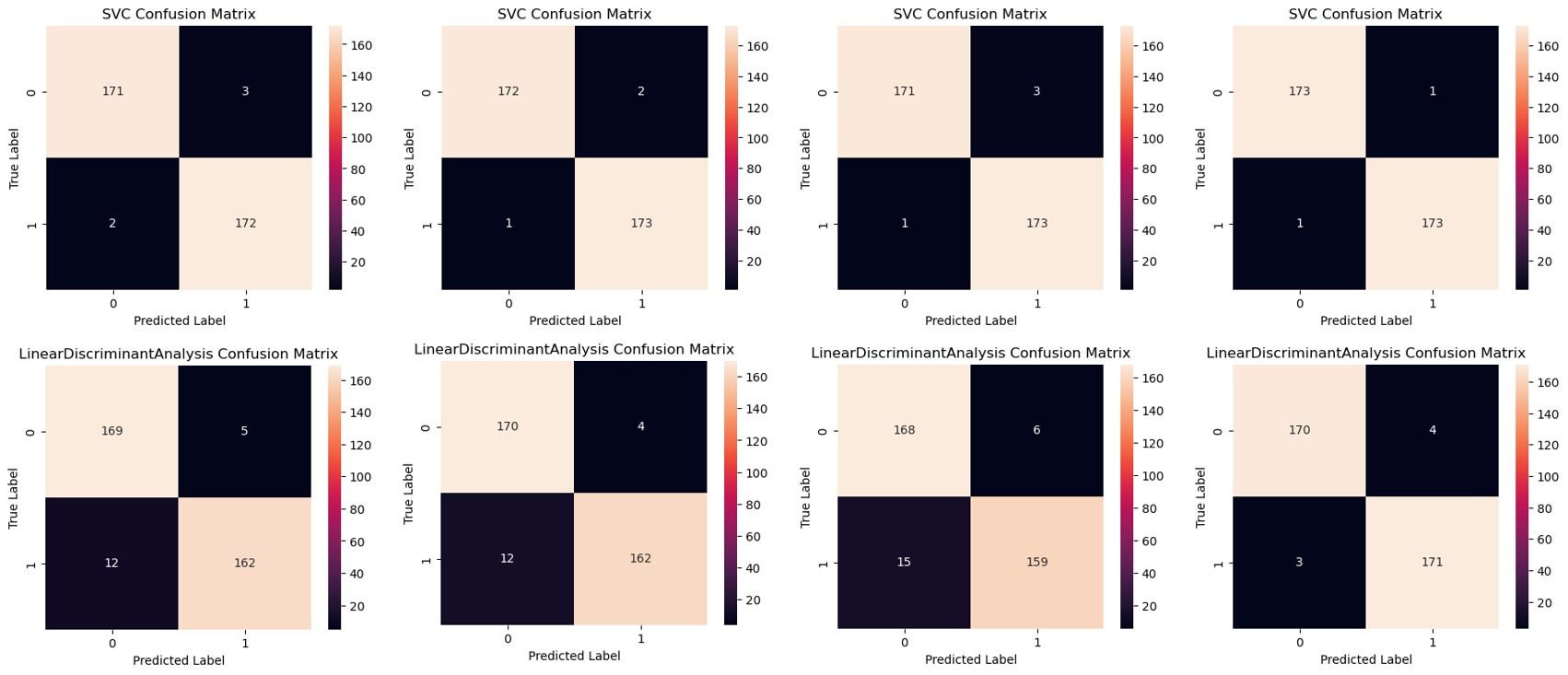}}

 \caption{This figure illustrates the 4x2 set of confusion matrices representing the performance of SVM and LDA models trained with various types of fNIRS data: oxy-Hb, deoxy-Hb, total-Hb, and a combination of all. The first row displays the confusion matrices of the SVM models: from left to right, the models are trained using oxy-Hb, deoxy-Hb, total-Hb, and a combination of all types of fNIRS data respectively. Similarly, the second row represents the LDA models trained with the same sequence of data types. Each matrix provides insights into the model's ability to correctly classify the instances, offering a visualization of true positive, true negative, false positive, and false negative counts.}
	\label{confusion_matrix}
\end{figure*}

\begin{figure*}[htbp]
	\centerline{\includegraphics[scale=0.62]{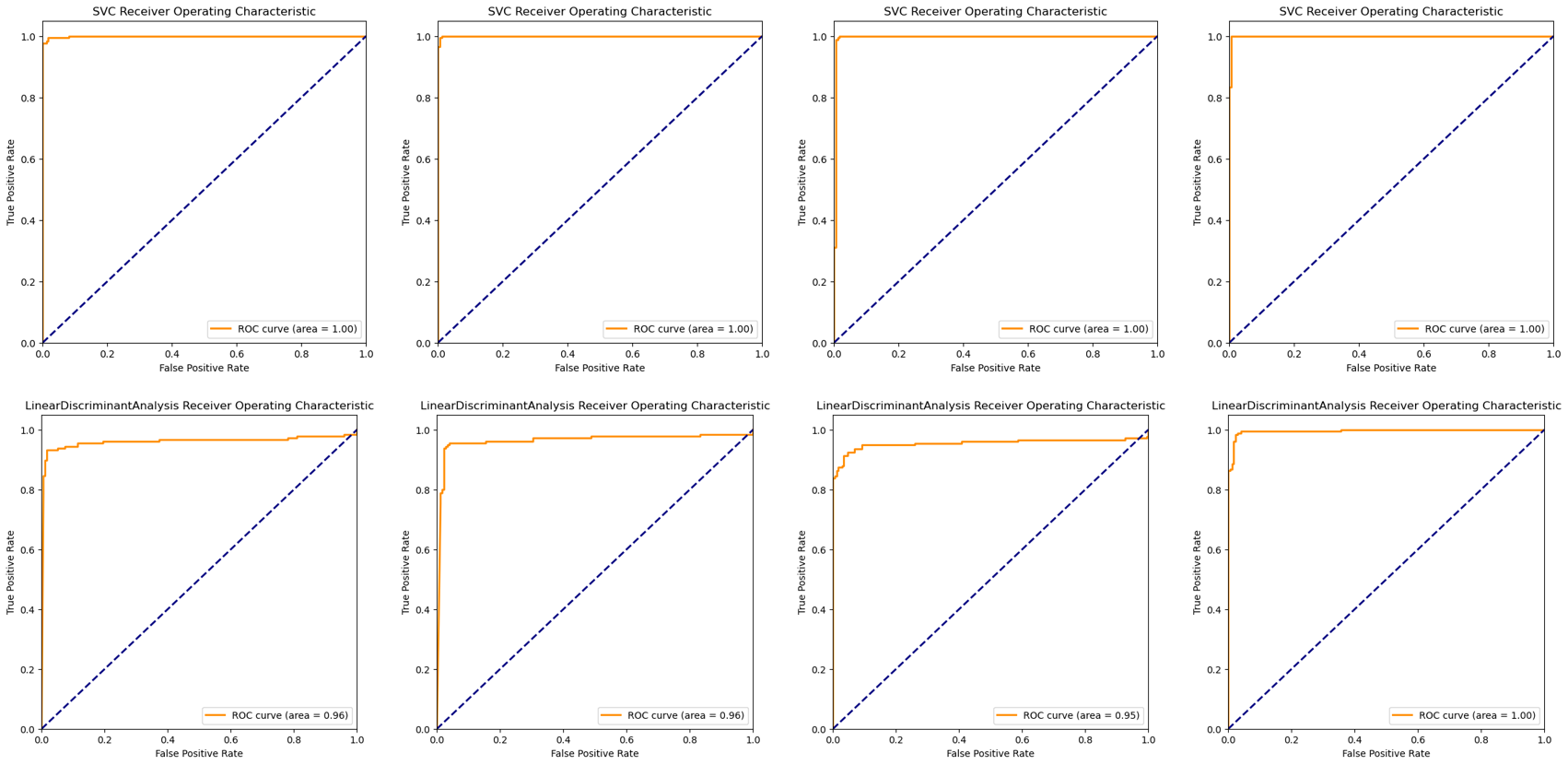}}
	\caption{This figure presents a 4x2 array of Receiver Operating Characteristic (ROC) curves, showcasing the discriminative ability of both SVM and LDA models when trained with different types of fNIRS data: oxy-Hb, deoxy-Hb, total-Hb, and a combination of all. In the first row, from left to right, the ROC curves represent SVM models trained respectively with oxy-Hb, deoxy-Hb, total-Hb, and combined data. The second row, with a similar arrangement, displays the ROC curves for LDA models. Each curve in the figure depicts the trade-off between the true positive rate (sensitivity) and false positive rate (1-specificity) at various thresholds, providing a comprehensive visualization of model performance across different data types. The area under each curve (AUC) serves as a single scalar value to summarize the model's ability to distinguish between the classes, with 1.0 representing perfect classification and 0.5 representing a model no better than random chance.}

	\label{ROC-curve}
\end{figure*}

Table~\ref{tab:Different alg result} shows the results of applying our model to nine machine learning algorithms. These include the classification correctness of oxy-­Hb, dexoy-­Hb, total-­Hb, and all metrics. Oxy-­Hb, a commonly used observable for fNIRS-based brain-computer interfaces\cite{ref34}, is represented by our feature extraction and normalization model on the Support Vector Machine (SVM) algorithm with the highest correct rate of 99.14\%. This is followed by applying our model to the Multi-Layer Perception (MLP) algorithm with a correct rate of 97.98\%. For most of the algorithms, the accuracy of our model is very high, which can also confirm the effectiveness of the model we designed. The same classifier as G.Bauernfein et al. is used is LDA and the accuracy of our model is improved by 28.09\% as compared to their model. The confusion matrix of SVM and LDA algorithms is shown in Fig.~\ref{confusion_matrix} and correspondingly, the ROC curve is shown in Fig.~\ref{ROC-curve}.

\begin{table}[htbp]
\centering
\caption{Subject-independent classification accuracies using various machine learning algorithms and our feature extraction method. }
\begin{tabular}{lllll}
\multicolumn{1}{l|}{\multirow{2}{*}{Algorithm}} & oxy-­Hb & deoxy-­Hb & total-­Hb & all   \\
\multicolumn{1}{l|}{}                           & \multicolumn{4}{l}{Acc.(\%)}            \\ \hline
\multicolumn{1}{l|}{SVM}                        & 99.14   & 99.14     & 98.00     & 99.43 \\
\multicolumn{1}{l|}{LR}                         & 96.84   & 98.00     & 95.97     & 99.13 \\
\multicolumn{1}{l|}{DT}                         & 79.91   & 88.50     & 83.93     & 86.20 \\
\multicolumn{1}{l|}{RF}                         & 95.13   & 96.84     & 92.52     & 96.85 \\
\multicolumn{1}{l|}{KN}                         & 83.06   & 74.41     & 78.14     & 80.77 \\
\multicolumn{1}{l|}{GNB}                        & 69.25   & 72.97     & 68.07     & 71.81 \\
\multicolumn{1}{l|}{LDA}                        & 95.69   & 96.55     & 93.67     & 98.00 \\
\multicolumn{1}{l|}{MLP}                        & 97.98   & 97.71     & 97.99     & 99.13 \\
\multicolumn{1}{l|}{SGD}                        & 96.56   & 94.85     & 88.77     & 96.56 \\ \hline
\multicolumn{5}{l}{\begin{tabular}[c]{@{}l@{}}$SVM$ Support Vector Machine, $LR$ Logistic Regression,\\ $DF$ Decision Tree, $RF$ Random Forst, $KN$ K Neighbors,\\ $GNB$ Gaussian NB, $LDA$ Linear Discriminant Analysis,\\ $MLP$ Multilayer Perceptron, $SGD$ Schotastic Gradient Descent\end{tabular}}
\end{tabular}
\label{tab:Different alg result}
\end{table}

\begin{table*}[htbp]
\centering
\caption{Performance metrics of the SVM and LDA algorithms trained on different types of fNIRS data: oxy-Hb, deoxy-Hb, total-Hb, and the combined dataset. The table enumerates Sensitivity (true positive rate), Specificity (true negative rate), Cohen's Kappa (a measure of classification accuracy corrected for chance), and the F1 Score (the harmonic mean of precision and recall) for both algorithms across the various data types. From the reported metrics, it can be observed how each algorithm's performance varies depending on the utilized fNIRS data. The comprehensive metrics provide insights into not only the models' correctness but also their reliability in differentiating between the classes, especially when it comes to unbalanced datasets.}

\label{tab:my-table}
\begin{tabular}{l|llll|llll|llll|llll}
\multirow{2}{*}{} & \multicolumn{4}{l|}{oxy-Hb}   & \multicolumn{4}{l|}{deoxy-Hb} & \multicolumn{4}{l|}{total-Hb} & \multicolumn{4}{l}{all}       \\ \cline{2-17} 
                  & TPR   & TNR   & K     & F1 (\%)    & TPR   & TNR   & K     & F1 (\%)   & TPR   & TNR   & K     & F1 (\%)   & TPR   & TNR   & K     & F1 (\%)    \\ \hline
SVM               & 98.85 & 98.28 & 97.13 & 98.56 & 99.43 & 98.85 & 98.28 & 99.14 & 99.43 & 98.28 & 97.70 & 98.84 & 99.43 & 99.43 & 98.85 & 99.43 \\
LDA               & 93.10 & 97.13 & 90.23 & 95.21 & 93.10 & 97.70 & 90.80 & 95.51 & 91.38 & 96.55 & 87.93 & 94.12 & 98.28 & 97.70 & 95.98 & 97.98
\end{tabular}
\end{table*}

In G. Bauernfeind's study, the Linear Discriminant Analysis (LDA) classification method was applied to eight subjects to compare the accuracy of the classification results, as shown in Table~\ref{tab:comparative_results}. The average accuracy of discrimination for the eight subjects was 63.28\% for Region of Interest 1 (ROI$_1$)and Region of Interest 2 (ROI$_2$), and 59.38\% for Region of Interest 3 (ROI$_3$). Obviously this is not a more desirable discrimination accuracy rate. This result is related to a number of factors, more than its failure to accurately extract data features. 
Although this approach cannot rule out individual differences between subjects, which may lead to salient experimental results for a particular subject, it can be used to validate the effectiveness of the research protocol.
\begin{table*}[htbp]
\centering
\caption{Comparative analysis of classification accuracies and [oxy-­Hb] patterns across three regions of interest (ROI$_1$, ROI$_2$, and ROI$_3$) between the experiments conducted by G. Bauernfeind et al. and our experiments. The table details the accuracy percentages, channel positions, Brodmann areas, and anatomical regions for each subject in both studies.}

\label{tab:comparative_results}
\begin{tabular}{llllllllllllll}
\multirow{3}{*}{Exp.} &
  \multirow{3}{*}{Sub.} &
  \multicolumn{4}{l}{ROI$_1$ {[}oxy-Hb{]}} &
  \multicolumn{4}{l}{ROI$_2$ {[}oxy-Hb{]}} &
  \multicolumn{4}{l}{ROI$_3$ {[}oxy-Hb{]}} \\ \cline{3-14} 
 &
   &
  \multirow{2}{*}{Acc.(\%)} &
  \multicolumn{3}{l}{Pos.} &
  \multirow{2}{*}{Acc.(\%)} &
  \multicolumn{3}{l}{Pos.} &
  \multirow{2}{*}{Acc.(\%)} &
  \multicolumn{3}{l}{Pos.} \\ \cline{4-6} \cline{8-10} \cline{12-14} 
 &      &       & Ch.    & BA & Anat. &       & Ch.    & BA & Anat. &       & Ch.    & BA & Anat. \\ \hline
\multirow{10}{*}{\begin{tabular}[c]{@{}l@{}}Experiments by G. Bauernfeind et al. \\ using LDA\cite{ref12}.\end{tabular}} &
  S1 &
  62.50 &
  46$^a$ &
  10 &
  SFG &
  56.25 &
  18$^c$ &
  9 &
  MFG &
  56.25 &
  24$^b$ &
  46 &
  MFG \\
 & S2   & 62.50 & 47$^a$ & 10 & MeFG  & 50.00 & 28$^c$ & 46 & MFG   & 62.50 & 24$^b$ & 46 & MFG   \\
 & S3   & 62.50 & 46$^a$ & 10 & SFG   & 50.00 & 29$^c$ & 9  & IFG   & 68.75 & 24$^b$ & 46 & MFG   \\
 & S4   & 81.25 & 48$^a$ & 10 & MFG   & 50.00 & 29$^c$ & 9  & IFG   & 62.50 & 24$^b$ & 46 & MFG   \\
 & S5   & 50.00 & 47$^a$ & 10 & MeFG  & 68.75 & 28$^c$ & 46 & MFG   & 56.25 & 24$^b$ & 46 & MFG   \\
 & S6   & 62.50 & 46$^a$ & 10 & SFG   & 50.00 & 28$^c$ & 46 & MFG   & 62.50 & 24$^b$ & 46 & MFG   \\
 & S7   & 62.50 & 47$^a$ & 10 & SFG   & 81.25 & 18$^c$ & 9  & MFG   & 62.50 & 23$^b$ & 46 & MFG   \\
 & S8   & 62.50 & 48$^a$ & 10 & MFG   & 68.75 & 28$^c$ & 46 & MFG   & 75.00 & 23$^b$ & 46 & MFG   \\
 & Mean & 63.28 &        &    &       & 59.38 &        &    &       & 63.28 &        &    &       \\
 & SD   & 8.48  &        &    &       & 12.05 &        &    &       & 6.19  &        &    &       \\ \hline
\multirow{10}{*}{\begin{tabular}[c]{@{}l@{}@{}}Our Experiments using LDA and our\\  feature extraction method\end{tabular}} &
  S1 & 94.25
   &
  46$^a$ &
  10 &
  SFG &
  89.00 &
  18$^c$ &
  9 &
  MFG &
  97.50 &
  24$^b$ &
  46 &
  MFG \\
 & S2   & 91.25 & 47$^a$ & 10 & MeFG  & 88.50 & 28$^c$ & 46 & MFG   & 94.75 & 24$^b$ & 46 & MFG   \\
 & S3   & 82.75 & 46$^a$ & 10 & SFG   & 86.00 & 29$^c$ & 9  & IFG   & 94.50 & 24$^b$ & 46 & MFG   \\
 & S4   & 98.00 & 48$^a$ & 10 & MFG   & 81.75 & 29$^c$ & 9  & IFG   & 100.00 & 24$^b$ & 46 & MFG   \\
 & S5   & 91.75 & 47$^a$ & 10 & MeFG  & 89.25 & 28$^c$ & 46 & MFG   & 87.50 & 24$^b$ & 46 & MFG   \\
 & S6   & 87.25 & 46$^a$ & 10 & SFG   & 87.50 & 28$^c$ & 46 & MFG   & 85.50 & 24$^b$ & 46 & MFG   \\
 & S7   & 83.25 & 47$^a$ & 10 & SFG   & 91.50 & 18$^c$ & 9  & MFG   & 91.50 & 23$^b$ & 46 & MFG   \\
 & S8   & 81.25 & 48$^a$ & 10 & MFG   & 98.00 & 28$^c$ & 46 & MFG   & 89.25 & 23$^b$ & 46 & MFG   \\
 & Mean & 88.72 &        &    &       & 88.94 &        &    &       & 92.56 &        &    &       \\
 & SD   & 6.05  &        &    &       & 4.65 &        &    &       & 5.01  &        &    &       \\  
  & SIA   & 90.23  &  46$^a$ & 10   &  SFG     & 86.20 &   18$^c$     &  9  &       MFG & 89.35  &  24$^b$      &  46  &   MFG    \\
 & SIA   & 87.64  &  47$^a$ & 10   &  MEFG     & 88.50 &   28$^c$     &  46  &       MFG & 89.10  &  23$^b$      &  46  &   MFG    \\
 & SIA   & 91.38  &  48$^a$ & 10   &  MFG     & 86.23 &   29$^c$     &  9  &  IFG &   &        &    &       \\ \hline
\multicolumn{14}{l}{$^a$ APFC, $^b$ r. DLPFC, $^c$ l. DLPFC}                                      \\
\multicolumn{14}{l}{$BA$ Brodmann area, $SFG$ superior frontal gyrus, $MFG$ middle frontal gyrus, $IFG$ inferior frontal gyrus, $MeFG$ medial frontal gyrus} \\
\multicolumn{14}{l}{$SIA$ Subject-independent Classification Accuracy}                        
\end{tabular}
\end{table*}

To further demonstrate the effectiveness of our model, we also used Linear Discriminant Analysis (LDA) for the same channels, ROI, and Brodmann areas as in G. Bauernfeind's study for eight subjects, and the classification accuracies obtained are shown in Table~\ref{tab:comparative_results}. The results show that the accuracy of the experimental results was significantly improved to varying degrees across subjects. The average discrimination accuracy for ROI$_1$ increased to 88.72\%, the average discrimination accuracy for ROI$_2$ increased to 88.84\%, and the average discrimination accuracy for ROI$_3$ incredibly increased to 92.56\%. This very well demonstrates the effectiveness of our well-designed feature extraction scheme. 

Furthermore, we conducted a statistical significance test based on Table IV. We employed the t-statistic to compare the mean accuracies of the two methods: ours and that proposed by G. Bauernfeind et al. A larger t-statistic implies a more distinct difference between the two groups. The p-value, on the other hand, denotes the probability of observing the given results, or more extreme results, assuming that no real difference exists between the two techniques. By convention, a p-value less than 0.05 is considered evidence of statistical significance.

It's worth noting that, for this analysis, we opted for Welch's t-test over the standard t-test. This decision was driven by our intent to make the test more robust to potential inequalities in the variances of the accuracies between the two methods. Welch's t-test is particularly recommended when there's a suspicion or evidence that the two data samples might have unequal variances, making it a more appropriate choice here.

\begin{table}[htbp]
\centering
\caption{Statistical significance analysis of differences in accuracy between the proposed method and the one by G. Bauernfeind et al. for ROIs.}
\label{tab:stat_results}
\begin{tabular}{l|ll}
ROI     & t-statistic & p-value \\ \hline
ROI$_1$ {[}oxy-Hb{]} & 6.91        & 0.00001 \\
ROI$_2$ {[}oxy-Hb{]} & 6.48        & 0.00011 \\
ROI$_3$ {[}oxy-Hb{]} & 10.39        & 8.58e-8 \\ \hline
\multicolumn{3}{l}{Note: p-value$<$0.5 represents statistically  } \\
\multicolumn{3}{l}{ significant difference. } 
\end{tabular}
\end{table}

From Table~\ref{tab:stat_results}, the observed t-statistics for ROI$_1$, ROI$_2$, and ROI$_3$ using Welch's t-test were 6.91, 6.48, and 10.39 respectively. The corresponding p-values were 0.00001, 0.00011, and 8.58e-8, all of which are substantially below the 0.05 threshold. These results offer compelling evidence of a significant accuracy enhancement when using our method in comparison to the approach of G. Bauernfeind et al.

In summation, our approach exhibits marked superiority in accuracy across all assessed ROIs, underscoring the merits of our model.

\subsection{New channel selection for fNIRS-based Online binary classification}
Using the data for machine learning after our proprietary feature extraction and normalization is valuable for fNIRS-based online binary classification. Previous fNIRS-based binary classification studies had a large number of channels to study\cite{ref21,ref22,ref23,ref24}, many redundant channels, and low accuracy. These large numbers of channels are not all channels that can be used to improve classification accuracy, and if higher accuracy can be achieved with fewer channels, this is certainly a study with applications. In order to improve the classification capability as well as the accuracy, and also to provide a more elaborate channel distribution scheme for fNIRS-based BCI device research.


\begin{figure*}[htbp]
	\centerline{\includegraphics[scale=0.55]{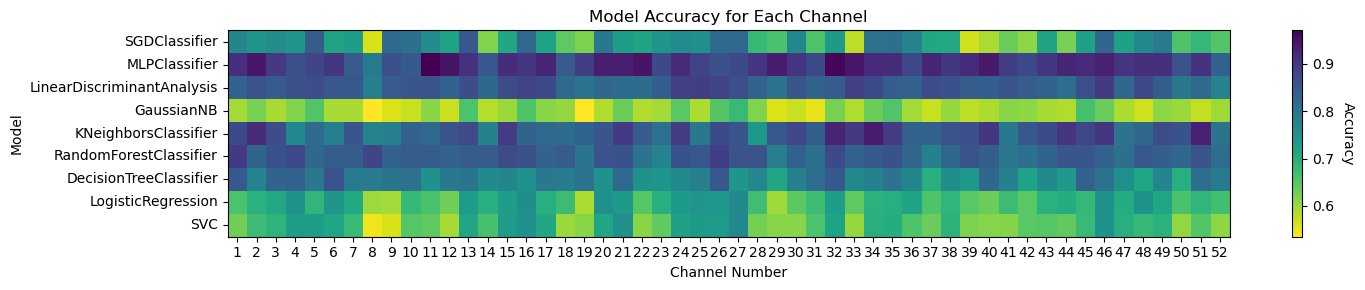}}
	\caption{Visualization of the subject-independent accuracy for various machine learning methods applied to fNIRS data channels, employing our feature extraction and normalization processing. In this heatmap, each column corresponds to one of the fNIRS data channels (ranging from 1 to 52), and each row represents a distinct machine learning method. The color intensity within the heatmap signifies the accuracy level, providing a comparative insight into the performance of different machine learning algorithms across various channels.}
	\label{acc1}
\end{figure*}

First, our model was used for different machine learning algorithms, and the accuracy was calculated for each of the 52 channels, and the results are shown in Fig.~\ref{acc1}. The horizontal coordinates are 1-52 channels, with different shades of color representing the correctness of different machine learning classifications. The results of the performance of the different algorithms can be visualized, and the multilayer perception algorithm shows a high and stable correct rate in each channel. This step serves to quickly eliminate some redundant channels with low correct rates, such as eight channels. The correct rate of single-channel classification can also be provided as a reference for comparison with multi-channel classification.

Then, based on the adjacency matrix constructed by the Pearson correlation coefficient, the 10 sets of two-by-two correlated channels with the lowest correlation were found. The correlation coefficients of the selected channels are lower than 0.4, which indicates that these key channels contain information that is more independent of each other and improves the accuracy of classification.

In Table~\ref{tab:top 10 channels}, we list the 10 channels with the highest accuracy single-channel classification result and their respective accuracy value. The highest accuracy is for channel 25, with a 92.76\% accuracy. Critically, we give the channels and their accuracy after channel combination. It is encouraging to find that there is a very significant improvement in the classification accuracy. Only for channel 26 and channel 43, the accuracy reaches 95.98\%. Compared with previous studies, the higher accuracy is obtained with a more compact number of channels, which reduces the complexity of data processing and algorithms, and greatly reduces the difficulty of device development.

\begin{table}[htbp]
\centering
\caption{Comparison of the top 10 individual channels and channel combinations in terms of their subject-independent accuracy. Each channel or combination is associated with a region (either APFC, r. DLPFC, or l. DLPFC) and for channel combinations, the Pearson correlation is provided to indicate the relationship between channels. This information serves to identify and understand the most influential channels and combinations in the fNIRS data.}
\begin{tabular}{l|ll|lll}
\multirow{2}{*}{No.} & \multicolumn{2}{l|}{Single Channel} & \multicolumn{3}{l}{Channel Combination} \\
                     & Ch.            & Acc.(\%)           & Ch.  & Acc.(\%)  & Pearson correlation  \\ \hline
1  & 25$^a$ & 92.76 & 26$^a$, 43$^b$ & 95.98 & 0.08118 \\
2  & 15$^a$ & 90.94 & 20$^c$, 26$^a$ & 95.40 & 0.16684 \\
3  & 16$^a$ & 90.16 & 23$^b$, 26$^a$ & 95.40 & 0.28594 \\
4  & 46$^a$ & 89.87 & 37$^a$, 43$^b$ & 95.12 & 0.03323 \\
5  & 36$^a$ & 89.63 & 26$^a$, 34$^b$ & 95.11 & 0.33568 \\
6  & 27$^a$ & 89.55 & 26$^a$, 51$^c$ & 94.84 & 0.20525 \\
7  & 26$^a$ & 88.98 & 26$^a$, 52$^c$ & 94.84 & 0.00340 \\
8  & 48$^a$ & 88.98 & 25$^a$, 52$^b$ & 94.84 & 0.33150 \\
9  & 12$^b$ & 88.67 & 37$^a$, 51$^b$ & 94.83 & 0.07331 \\
10 & 34$^b$ & 88.64 & 37$^a$, 52$^b$ & 94.55 & 0.10958\\ \hline
\multicolumn{6}{l}{$^a$ APFC, $^b$ r. DLPFC, $^c$ l. DLPFC}                                      \\
\end{tabular}
\label{tab:top 10 channels}
\end{table}

\begin{figure}[htbp]
	\centerline{\includegraphics[scale=0.32]{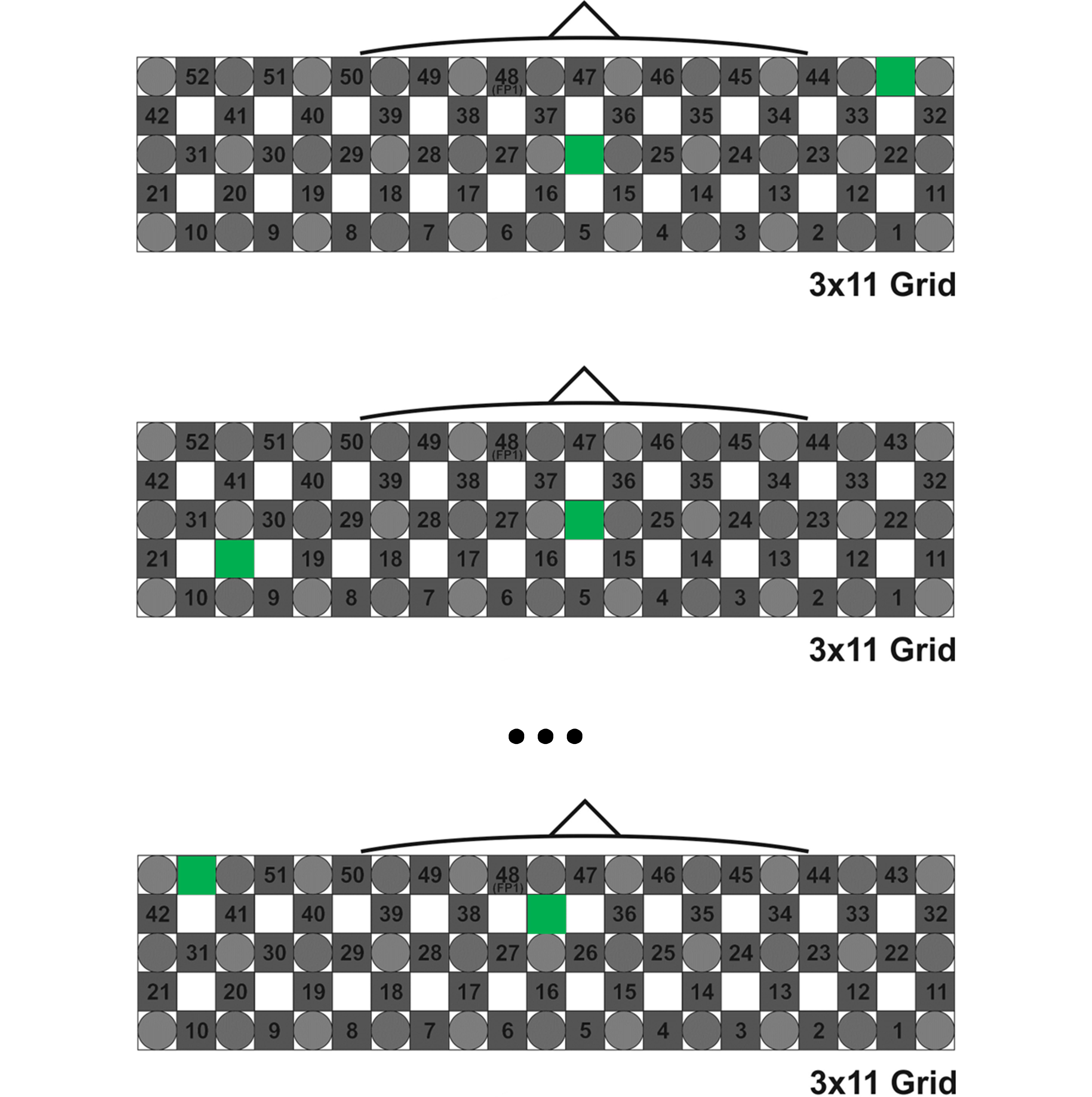}}
	\caption{The combination of channels that have been selected by our scheme is highlighted in green in the diagram, while the other channels are shown in gray to highlight their positional information.}
	\label{ccf}
\end{figure}

\section{Discussion and Future Work}
In this work, we aim to address the dual challenges associated with online classification of fNIRS data and achieving subject-independent accuracy. To tackle these challenges, we propose a feature extraction scheme that is hemodynamically compliant for fNIRS data. This scheme involves constructing feature matrices comprising statistical features, time-domain, frequency-domain, and principal component features.



Our approach presents a paradigm for the preliminary development process of fNIRS-based BCIs tailored for binary classification tasks, maintaining high subject-independent accuracy while reducing the number of transmitters and receivers. Theoretically, our methodology holds potential for extension to any binary classification challenge within fNIRS-based frameworks. This investigation presents a foundational blueprint for binary classification in fNIRS-based BCI platforms, holding promise for addressing diverse binary classification dilemmas within fNIRS-centric frameworks.

\bibliographystyle{IEEEtran}
\bibliography{reference}

\end{document}